\documentclass[12pt]{article}

\newcommand{\be}{\begin{equation}}
\newcommand{\ee}{\end{equation}}
\newcommand{\bey}{\begin{eqnarray}}
\newcommand{\eey}{\end{eqnarray}}
\newcommand{\non}{\nonumber\\}

\renewcommand{\vec}[1]{\mbox{\boldmath{$\scriptstyle #1$}}}
\newcommand{\mbf}[1]{\mbox{\boldmath{$#1$}}}

\renewcommand{\int}{\intop\limits}
\title{\begin{flushright}
{\normalsize NUC-MINN-97/5-T\\
June 1997 \\}
\end{flushright}
\vspace*{0.3in}
{\bf Fluctuation and Dissipation in Classical Many-Particle Systems}}

\author{{\bf L. P. Csernai}
\\[1.5ex]
{\it Section for Theoretical Physics, Department of Physics}\\
{\it University of Bergen, Allegaten 55, 5007 Bergen, Norway}\\
[1.5ex]
{\bf S. Jeon, J. I. Kapusta}\\[1.5ex]
{\it School of Physics and Astronomy, University of Minnesota}\\
{\it 116 Church Street SE, Minneapolis, Minnesota 55455, USA}}

\date{}

\begin{document}

\maketitle

\begin{abstract}
Coarse-grained Langevin-type effective field equations
are derived for classical
systems of particles.  These equations include the effects of
thermal fluctuation and dissipation which may arise from coupling
to an external bath, as in the Brownian motion of a single particle,
or which may arise from statistical fluctuations in small parts
of an isolated many-particle system, as in sound waves.  These equations
may provide some guidance for the analysis of mesoscopic or microscopic
molecular systems, or for systems of hundreds to thousands of
subatomic particles produced in high energy nuclear collisions.
\end{abstract}

\vfill
\noindent
csernai@fi.uib.no\\
jeon@nucth1.spa.umn.edu\\
kapusta@physics.spa.umn.edu\\

\newpage

\section{Introduction}

The theoretical description of Brownian motion of a classical particle
is well known and widely discussed; it is textbook material
\cite{Reif65}.  It is used in numerous practical applications,
such as evaluating reaction rates at finite temperature in systems
where thermal fluctuations are of vital importance
\cite{kr40,zw59,vi76,cn82}.  However, for continuous classical
systems thermal fluctuations are not so frequently
discussed, and the actual evaluation of fluctuating forces and dissipative
coefficients for dense, interacting, classical, continuous systems is
not so well known.

We were originally motivated to this study by the physics of nuclear
collisions at high energy.  In these collisions many subatomic particles
are produced, mostly pions which are the main carrier of the nuclear
force.  The collision may be viewed as a miniature Big Bang where
soon after impact a large amount of the initial translational energy
is put into particle creation and entropy production.  This system
can be roughly characterized by a temperature.  As time goes on, the
system expands and cools.  Eventually collisions become so infrequent
that thermal equilibrium is lost and the particles free-stream to
infinity and are detected.  The properties of a system of pions,
numbering in the hundreds or thousands, at high temperature may
have very interesting properties.  For example, pion fields transform
under an internal symmetry group rather analogous to spins in
a magnet.  During the early stage of the expansion, when the
temperature is high, the field may collectively point in a
direction different than it does in the surrounding vacuum.
This is referred to as a disoriented chiral condensate (DCC)
\cite{Wilczek}.  One would like to have coarse-grained field
equations to describe the fluctuation and dissipation
of DCC domains and their inevitable coalescence and evolution
into the sourrounding vacuum \cite{somedcc}.  A good description
is lacking.

Our goal here is to develop some understanding of the DCC problem by
considering a collection of classical particles undergoing Brownian motion
and generalizations of such.  This might appear to be a simple problem
but it seems not to have been discussed in the literature.  In particular
there are subtle issues relating to the nature of what one considers
the heat bath.  For a single particle it is relatively straightforward,
for a collection of particles it is not.  In one limit, the particles
of interest may each be coupled to an external heat bath but they
may also interact with each other via forces that are more slowly
varying than the ones operative between the particles and the heat
bath.  In this case Langevin equations of motion can be obtained
for time scales that are short compared to the interparticle
interaction times but long compared to the interaction times with
the heat bath.  In another limit, there is no external heat bath;
one averages over a macroscopically small but microscopically large
number of neighboring particles and seeks a Langevin equation to describe
the motion of these subsets of particles over times long compared
to the force fluctuation times between subsets.  In the real world
there may be a continuum of interesting problems lying between
these two extremes.

In the following sections we will analyze the two limiting cases
outlined above.  In both cases we will do coarse-graining to
obtain an effective field equation of motion, in one and three dimensions.

\section{External Heat Bath}

\subsection{Recollection of simple Brownian motion}

This case is discussed in many textbooks, such as \cite{Reif65},
and often repeated in the literature, such as \cite{cortes}.
For a particle of mass $m$ connected to a heat bath and
moving under the influence of external potential fields, for example
a small object suspended in a fluid or gas in a gravitational
field, and moving in one
dimension, the generalized Langevin equation is
\be
m\frac{dv(t)}{dt} = G(t) -\int_{-\infty}^t dt' \, K(t{-}t')
v(t') + F'(t) \, .
\ee
Here $v(t)$ is the velocity of the particle at time $t$.
The force due to external fields is labeled by $G$.
The force due to the heat bath has been separated
into two parts:  $F = \overline{F} + F'$, where
$F'$ represents the rapidly varying, random, part of the force
whose average value is zero and $\overline{F}$ represents the
slowly varying part whose average value is not necessarily zero.
The separation of these
two components depends on the coarse-graining time chosen.
To be a useful coarse-graining, this time must be large
compared to the characteristic correlation time $\tau_{\rm cor}$
of the force, but must be small enough on an observational time
scale to record the desired coarse-grained trajectory of the particle.
For example, if the particle under consideration is immersed
in a gas of much lighter particles, $\tau_{\rm cor}$ is the
average time between collisions with the gas particles
which define the heat bath.  The friction
kernel is
\be
K(s) = \beta \langle F'(0) F'(s) \rangle
\ee
where $\beta$ is the inverse temperature of the heat bath
(we use units with $k_B = 1$ throughout) and the averaging
is carried out with respect to the heat bath.  This is referred to
as a generalized Langevin equation; the normal Langevin equation
is obtained in the approximation that $K(s) = 2\alpha \delta(s)$.
This may be considered as the limit of the somewhat more general
case where
\be
K(s) = \frac{\alpha}{\tau_{\rm cor}}
\exp\left(-|s|/\tau_{\rm cor}\right) \, .
\ee
Simple Brownian motion in more than one dimension is obvious.

\subsection{Coupled Brownian particles}

\begin{figure}[htb]
\setlength{\unitlength}{0.5mm}
\begin{center}
\begin{picture}(300,190)(0,0)
%
%    Bottom unit fig. 1-2
%
\thicklines
%
% tails length 5
\put(  5,100){\line( 1, 0){ 5}}
\put(290,100){\line( 1, 0){ 5}}
% 5 box labels
\put( 17,95){\LARGE $B_{\mu-2}$}
\put( 77,95){\LARGE $B_{\mu-1}$}
\put(142,95){\LARGE $B_{\mu}$}
\put(197,95){\LARGE $B_{\mu+1}$}
\put(257,95){\LARGE $B_{\mu+2}$}
%
% box 40x30  (tot. width 40) 5 of these
\multiput(0,0)(60,0){5}{
\put(10,85){\line(1,0){40}} \put(10,115){\line(1,0){40}}
\put(10,85){\line(0,1){30}} \put(50, 85){\line(0,1){30}}
}
%
% spring 12x18 with tails 4-4 (tot. width 20) 4 of these
\multiput(0,0)(60,0){4}{
\put(50,  100){\line(1, 0){4.0}} \put(54.0,100){\line(1,6){1.5}}
\put(55.5,109){\line(1,-6){3.0}} \put(58.5, 91){\line(1,6){3.0}}
\put(61.5,109){\line(1,-6){3.0}} \put(64.5, 91){\line(1,6){1.5}}
\put(66.0,100){\line(1, 0){4.0}}
}
%
%
%       Top Unit Fig. 1
%
% box "Heat Bath" H: 280x40  (tot. width 40)
\thicklines
\put(10,140){\line(1,0){280}} \put( 10,180){\line(1,0){280}}
\put(10,140){\line(0,1){ 40}} \put(290,140){\line(0,1){ 40}}
%
%  interaction lines
         \thinlines           \put( 30,115){\line(0,1){25}}
\put( 90,115){\line(0,1){25}} \put(150,115){\line(0,1){25}}
\put(210,115){\line(0,1){25}} \put(270,115){\line(0,1){25}}
%  box label
\put(110,155){\LARGE Heat Bath $H$}
\end{picture}
\end{center}
\vspace*{-4.5cm}
\caption[]{
A series of Brownian particles, $B_{\mu{-}1}$, $B_{\mu}$, $B_{\mu{+}1}$,
..., interacting with
a thermal heat bath $H$ with random thermal forces
$F_{\mu}(t)$ (thin lines)
and with their nearest neighbours via conservative forces (heavy lines).
}
\label{fig:f_heat}
\setlength{\unitlength}{1mm}
\end{figure}
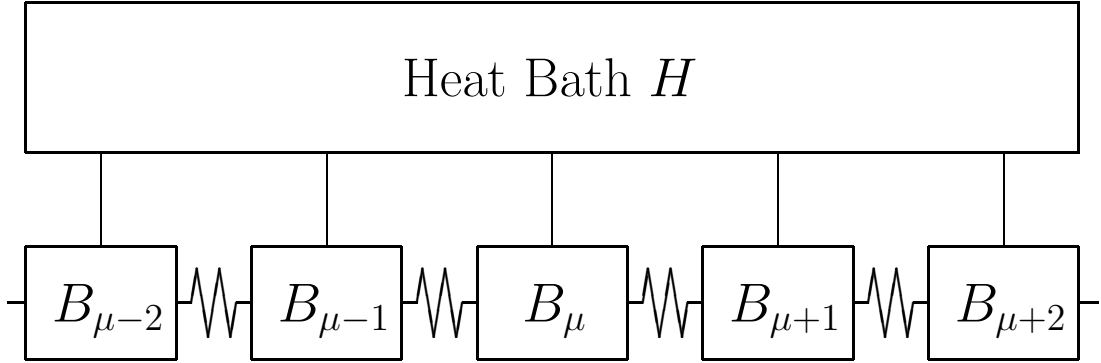

Consider a set of Brownian particles referred to as $B_{\mu}$ which can
move in one dimension and which are in thermal
interaction with a heat bath referred to as $H$; see
Fig. \ref{fig:f_heat}.
%\pagestyle{myheadings}
%\markboth{\rm Csernai, L.P., Kapusta, J.I., and J. Sangyong}%
%{\rm Thermal Fluctuations in Field Theory}
The thermal interaction is mediated by a fluctuating force $F_{\mu}(t)$
which has a mean period of $\tau_{\rm cor}$. This time, $\tau_{\rm cor}$,
characterizes
the relaxation time needed for $H$ to reestablish its equilibrium
configuration if it is perturbed by some sudden small change. We assume
that the conservative force connecting Brownian particles is slowly varying
compared to $\tau_{\rm cor}$; we will study the change of position and
velocity of the Brownian particles assuming that changes are small in a time
interval $\tau > \tau_{\rm cor}$.

Suppose that at time $t$ all $B_{\mu}$ are in thermal equilibrium with $H$.
Consider a set of macroscopically similar systems
forming an ensemble. Then the ensemble average of the fluctuating forces
vanishes at this moment.
\be
\langle F_{\mu}(t) \rangle_0 = 0
\ee
The subscript indicates that the average is taken in thermal
equilibrium.  Due to a change in the position or velocity of some $B_{\mu}$
the force $F_{\mu}$ may change and our
system, both $B_{\mu}$ and $H$, may deviate from thermal
equilibrium.  We have to evaluate this deviation.

Apart from the fluctuating random force $F_{\mu}(t)$ there is another
slowly varying, not fluctuating, force $G_{\mu}(t)$.
The forces connecting the Brownian particles are of this
type. We will be interested in finding the corresponding slowly
varying part of the velocity of each $B_{\mu}$. Integrating the
equation of motion,
$m \dot{v}_{\mu} = G_{\mu}(t) + F_{\mu}(t)$,
for a short but macroscopic period of time
$\tau$ we get, in a particular microscopic configuration,
\begin{equation}
m \left[v_{\mu}(t{+}\tau) - v_{\mu}(t)\right] = G_{\mu} \tau +
 \int_t^{t{+}\tau} F_{\mu}(t') dt'  \ .
\label{eom-0}
\end{equation}
Here we have taken into consideration that $G_{\mu}$ can be
considered constant during the short period of time $\tau$.
The microscopic configurations are not under our control;
we know only the ensemble average of similarly prepared systems. Taking
the ensemble average of both sides of the above equation of motion
we obtain
\be
m \left\langle v_{\mu}(t{+}\tau) - v_{\mu}(t)\right\rangle =
G_{\mu} \tau +
       \int_t^{t{+}\tau} \langle F_{\mu}(t') \rangle dt'  \ .
\ee
In general $\langle F_{\mu}(t') \rangle$ is not independent of the
motion of $B_{\mu}$, otherwise it would be always the same as the mean
value $\langle F_{\mu}(t) \rangle_0 = 0$ in thermal equilibrium.
We will evaluate the changes in
$ \langle F_{\mu}(t') \rangle$ following the lines
of \cite{Reif65} (sect.15.7).

Let us consider the change during the time interval from $t$ to $t{+}\tau'$.
The velocity of $B_{\mu}$ changes from
$v_{\mu}(t)$ to $v_{\mu}(t{+}\tau')$.  The
motion of this particle affects its environment.  If $\tau'$ is small
enough the mean force $\langle F_{\mu}\rangle$ changes but still
depends on its earlier value at $t$.
If $\tau'$ exceeds $\tau_{\rm cor}$, the heat bath
will reestablish its thermal equilibrium and will be found with equal
probability in any of its $\Omega$ accessible states. Since the energy of
the $B_{\mu}$ change the energy of the heat bath changes as well.  The
total energy of the heat bath changes by
\be
\Delta E(\tau') =  ... + \Delta E_{\mu-1} (\tau')
                        + \Delta E_{\mu} (\tau')
                        + \Delta E_{\mu{+}1} (\tau')  + ... ,
\ee
where $\Delta E_{\mu}(\tau')$ is the energy {\it given} to the
heat reservoir $H$ {\it by} the particle $B_{\mu}$.
Since we consider small time increments only the local environment of
$B_{\mu}$ is relevant.  The number of other Brownian particles
influencing the heat
bath is not infinite but extends to some distance comparable to the
mean free path or the spatial correlation length.
The number of states available to $H$
changes from $\Omega(E_0)$ to $\Omega(E_0 + ... +
\Delta E_{\mu{-}1} + \Delta E_{\mu} + \Delta E_{\mu{+}1}  + ... )$.
This change of energy of the heat bath will influence the Brownian
particles connected to it and modify the populations of microstates $r$
in the ensemble corresponding to $B_{\mu}$.  We will use this information
to estimate the change of the fluctuating thermal force acting on $B_{\mu}$.

The equilibrium probability  of the occurrence of a given microstate $r$
for $B_{\mu}$ is proportional to the corresponding number of states
available to the heat bath.  Here we assume that the total number
of microstates of all Brownian particles $B_{\mu{-}1}$, $B_{\mu}$,
$B_{\mu{+}1}$, ...  together is negligibly small compared to the
number of microstates of the heat reservoir.
Now we can compare the probability distribution over the microstates
$W_{r\mu}$ of $B_{\mu}$ at time $t$ and $\tau'$ later.
\begin{eqnarray}
\frac{W_{r\mu}(t{+}\tau')}{W_{r\mu}(t)} &=&
\frac{W_{r\mu}(t{+}\tau')}{W_{r\mu}^{(0)}} =
\frac{\Omega(E_0+ ... +\Delta E_{\mu{-}1} +\Delta E_{\mu}
 +\Delta E_{\mu{+}1}+ ... )}{\Omega(E_0)} \nonumber \\
&=& \exp\left[\beta( ... +\Delta E_{\mu{-}1} +\Delta E_{\mu}
+\Delta E_{\mu{+}1}+ ...)\right]
\end{eqnarray}
Here $\beta \equiv \partial \ln \Omega / \partial E$ is the inverse
temperature of the heat bath, assumed to be constant. This means that
if more energy is made available to the heat bath the probability to
populate a particular microstate $r$ of $B_{\mu}$ increases.
This is true even if that particular $B_{\mu}$ takes energy from
the heat bath as long as its neighbours add more.

Now we can estimate how population probabilities change with time.
\begin{eqnarray}
W_{r\mu}(t{+}\tau') &=& W_{r\mu}^{(0)}
\exp[\beta( ... +\Delta E_{\mu{-}1} +\Delta E_{\mu}
 +\Delta E_{\mu{+}1}+ ... )] \nonumber \\
&\approx& W_{r\mu}^{(0)}
[ 1+ \beta( ... +\Delta E_{\mu{-}1} +\Delta E_{\mu} +\Delta E_{\mu{+}1}+ ... )]
\end{eqnarray}
We can also evaluate how the ensemble average of the fluctuation force
changes during this time interval.
\begin{eqnarray}
\langle F_{\mu} (t{+}\tau') \rangle &\equiv&  \sum_r
W_{r\mu}(t{+}\tau') F_{r\mu} \nonumber \\ &\approx&  \sum_r W_{r\mu}^{(0)}
[ 1+ \beta( ... +\Delta E_{\mu{-}1} +\Delta E_{\mu}
 +\Delta E_{\mu{+}1}+ ... )] F_{r\mu} \nonumber \\
&=& \langle [ 1+ \beta( ... +\Delta E_{\mu{-}1} +\Delta E_{\mu}
 +\Delta E_{\mu{+}1}+ ... )] \ F_{\mu} \rangle_0
\end{eqnarray}
Since the ensemble average of the fluctuating force vanishes for the
thermal equilibrium distribution $W_{r\mu}^{(0)}$
the above expression for the fluctuation force reduces to
\be
\langle F_{\mu} (t{+}\tau') \rangle =
\beta  \langle ( ... +\Delta E_{\mu{-}1} +\Delta E_{\mu}
 +\Delta E_{\mu{+}1}+ ... ) F_{\mu} \rangle_0 \, .
\ee
In this estimate the change of the energy of the heat bath is still
undefined. This energy is, however, simply the negative of the work done
by the fluctuating force on the Brownian particle $B_{\nu}$.
\be
\Delta E_{\nu} = - \int_t^{t+\tau'} F_{\nu}(t'') \ v_{\nu}(t'') \ dt''
\ee
Inserting this expression into that for the fluctuating force
we obtain
\begin{equation}
\langle F_{\mu} (t{+}\tau') \rangle =
- \beta
\int_t^{t+\tau'}  dt'' \
\langle F_{\mu}(t{+}\tau') F_{\nu}(t'') \rangle_0  \ v_{\nu}(t'') \, .
\label{dej}
\end{equation}
We use the summation convention where a repeated index is summed over.
In this case, the sum on $\nu$ runs over the neighbors of $B_{\mu}$;
it is cut off by
the range of the forces and by the finite time interval.
There is no need to do an ensemble averaging over the velocities
because they are much more slowly varying than the fluctuating
forces.  Thus we have estimated the expectation value of the force
$F_{\mu}$ for an
ensemble, weakly deviating from a thermal equilibrium,
via expectation values obtained in thermal equilibrium.
This is, of course, just linear response theory.

Now the equation of motion for the Brownian particle $B_{\mu}$
can be cast in the form
\be
m \langle v_{\mu}(t{+}\tau) - v_{\mu}(t) \rangle = G_{\mu} \tau
- \beta  \int_t^{t{+}\tau} dt' \int^{t'-t}_0  ds \
\langle  F_{\mu}(t') F_{\nu}(t'{-}s) \rangle_0 \ v_{\nu}(t'{-}s) \, .
\label{ptmeom}
\ee
The ensemble averaging in the expression for the interparticle
force may be dropped because the coordinates
change much more slowly than the velocities, which are
again much more slowly varying than the fluctuating forces.

It is straightforward to perform the above derivation in 3 dimensions. The
forces $F_{\mu}$ and velocities $v_{\mu}$ become 3-dimensional
vectors $F^i_{\vec{\mu}}$ and $v^i_{\vec{\mu}}$ with spatial indices labeled
by Roman letters and where the bold Greek indices now label the
position.  The work done by particle $B_{\vec{\nu}}$ will contain the
scalar product $v_{\vec{\nu}}^j F_{\vec{\nu}}^j$ of its velocity and the
random force acting on it. Eqn. (\ref{ptmeom}) takes the form
\be
m \langle v^i_{\vec{\mu}}(t{+}\tau) - v^i_{\vec{\mu}}(t) \rangle =
G^i_{\vec{\mu}} \tau
- \beta \int_t^{t{+}\tau} dt' \int^{t'{-}t}_0 ds \
\langle F^i_{\vec{\mu}}(t') F^j_{\vec{\nu}}(t'{-}s) \rangle_0
 \ v^j_{\vec{\nu}}(t'{-}s) \, .
\label{ptmeom3}
\ee

The friction kernel is proportional to the correlation function
of the random fluctuating force.
\be
K^{ij}_{\vec{\mu\nu}}(s) \equiv \beta
\langle F^i_{\vec{\mu}}(t) F^j_{\vec{\nu}}(t{-}s) \rangle_0
\ee
Here it is assumed that this function is time translation invariant,
as is normally the case in thermal equilibrium.  This assumption
can be relaxed.  We will discuss the properies of time
translationally invariant correlation functions a little later.

Now it is useful to separate the fluctuating force into two
components.  The first component is just the average value
of the fluctuating force in the slightly out of equilibrium
ensemble.
\be
\langle F_{\vec{\mu}}^i(t) \rangle = -
\int_t^{t+\tau} \frac{dt'}{\tau} \int^{t{-}t'}_0 ds \
K^{ij}_{\vec{\mu\nu}}(s) \ v_{\vec{\nu}}^j(t'{-}s)
\ee
It leads to a damping of the velocity.  Note that the right
side of this expression involves averages defined with respect to
the unperturbed thermal ensemble.
We note that the integrand is appreciable only when $s \ll \tau$
because the coarse graining time $\tau$ was chosen to be much
bigger than the correlation time of the random forces.
The velocity is slowly varying and so it may be evaluated with
$t$ replacing $t'$.  (This is just the first term in a Taylor
series expansion: $v(t'-s) = v(t-s) + (t'-t)\dot{v}(t-s)
 + \cdots$).  The upper limit of the $s$ integral may then
be sent to infinity yielding the approximation
\be
\langle F_{\vec{\mu}}^i(t) \rangle = -
\int^{\infty}_0 ds \ K^{ij}_{\vec{\mu\nu}}(s) \ v_{\vec{\nu}}^j(t{-}s)
\ee
The nonlocality reflects the time delay between the motion
of the particles and the responding force.

The second component is the most rapidly fluctuating
part and is defined by its zero average with respect to the
actual out of equilibrium ensemble.
\be
F^{\prime \, i}_{\vec{\mu}}(t) \equiv F_{\vec{\mu}}^i(t)
 - \langle  F_{\vec{\mu}}^i(t) \rangle
\ee
The equation of motion including both the drift and the fluctuating
parts of the velocity is
\begin{equation}
m\frac{dv_{\vec{\mu}}^i(t)}{dt} = G_{\vec{\mu}}^i
- \int^{\infty}_0 ds \ K^{ij}_{\vec{\mu\nu}}(s) \ v_{\vec{\nu}}^j(t{-}s)
+ F_{\vec{\mu}}^{\prime \, i}(t) \, .
\label{langevin1}
\end{equation}

In this analysis the stochastic forces of the external heat bath
act on every Brownian particle the same way. The correlation between
two forces, $K^{ij}_{\vec{\mu\nu}}(s)$, under normal circumstances is
expected to decrease with increasing distance between
Brownian particles $B_{\vec{\mu}}$ and $B_{\vec{\nu}}$,
and with the time difference $s$.
The summation over the neighbours $\vec{\nu}$ of $B_{\vec{\mu}}$ extends to
infinity in all directions; however, the contribution of more and more
distant neighbours is expected to be rapidly decreasing.

\subsection{Continuum limit}

Rather than attempting to solve the equations of motion for a macroscopic
number of particles it is oftentimes useful to approximate them
by a continuous medium.  This is effectively coarse graining, and is
accurate so long as the length scales of interest are large enough
and the time scales of interest long enough.  With this in mind, let
us replace the discrete particle labels $\mbf{\mu}$ and $\mbf{\nu}$ with
continuous position variables ${\bf x}$ and ${\bf y}$.
Displacement of the particles in the gas will be denoted by
${\eta}^i({\bf x},t)$.  The sum over particle index
is replaced by an integral over position.
Divide both sides of eq. (\ref{langevin1}) by the average
volume $v_0$ per particle.  The mass density is
$\rho \equiv m/v_0$.  Forces per particle then become forces
per unit volume and are denoted by a lower case letter.
The Langevin equation then becomes
\be
\rho \, \ddot{\eta}^i({\bf x},t)
= g^i({\bf x},t) - \int d^3y \int_0^{\infty} ds \
k^{ij}({\bf y},s) \ \dot{\eta}^j({\bf x}-{\bf y},t-s)
+f^{\prime \, i}({\bf x},t) \, .
\ee
Here
\be
k^{ij}({\bf y},s) = \beta \langle f^i({\bf x},t)
f^j({\bf x}-{\bf y},t-s) \rangle_0
\ee
is the correlation function for the force densities of the
heat bath.

If the medium is isotropic then the friction kernel must have
the tensorial structure
\be
k^{ij}({\bf y},s) = k_{\rm L}(y,s) \hat{y}^i \hat{y}^j +
k_{\rm T}(y,s) \left( \delta^{ij} - \hat{y}^i \hat{y}^j \right)
\ee
where $y = |{\bf y}|$, and $k_{\rm L}$ and $k_{\rm T}$ are
longitudinal and transverse correlation functions.
When we are interested in lengths
greater than those characterizing the friction kernel we
can evaluate the velocity at position ${\bf x}$ and take
it past the $y$ integration.  (More generally, $\dot{\eta}$
would be expanded in a Taylor series about ${\bf y} = 0$.)
Carrying out the averaging over angles gives
\be
\rho \, \ddot{\mbox{\boldmath$\eta$}}({\bf x},t)
= {\bf g}({\bf x},t) -  \int_0^{\infty} ds \
\dot{\mbox{\boldmath$\eta$}}({\bf x},t-s)
\int d^3y \
\left( \frac{1}{3}k_{\rm L}(y,s) + \frac{2}{3} k_{\rm T}(y,s) \right)
+{\bf f}^{\prime \, i}({\bf x},t) \, .
\ee
Usually it happens that correlations fall off exponentially.
Then the friction kernel may be parametrized by
correlation times, correlation lengths, and strengths.
\begin{eqnarray}
k_{\rm L}(y,s) &=& \beta \langle f^2_{\rm L} \rangle
\exp\left(-s/\tau_{\rm L} -y/\lambda_{\rm L} \right) \nonumber \\
k_{\rm T}(y,s) &=& \beta \langle f^2_{\rm T} \rangle
\exp\left(-s/\tau_{\rm T} -y/\lambda_{\rm T} \right)
\end{eqnarray}
When our interest is on times greater than those
characterizing the friction kernel and when the actual displacement
velocity is slowly varying on those scales we can take the
velocity outside the $s$ integration to obtain
\be
\rho \, \ddot{\mbf\eta}({\bf x},t)
= {\bf g}({\bf x},t) -  \rho \gamma \dot{\mbf\eta}({\bf x},t)
+{\bf f}'{\bf x},t) \, ,
\ee
where the damping constant is
$\gamma = \frac{1}{3}\gamma_{\rm L} + \frac{2}{3}\gamma_{\rm T}$ with
\be
\gamma_{\rm L} = \frac{\langle f^2_{\rm L} \rangle}{\rho T}
\int_0^{\infty} ds \int d^3y \
\exp\left(-s/\tau_{\rm L} -y/\lambda_{\rm L} \right)
= 4\pi \Gamma (3) \frac{\langle f^2_{\rm L} \rangle
\tau_{\rm L} \lambda_{\rm L}^3}{\rho T}
\ee
and with a similar expression for $\gamma_{\rm T}$.
This is a manifestation of the fluctuation dissipation theorem.

The interesting feature of this Langevin field equation is that
the damping term is linear in the displacement velocity.  This
is the form which is normally used in phenomenological settings.
However, this is certainly not the most general Langevin field
equation for the displacement, as may be observed from the
truncation of the Taylor expansion in ${\bf y}$, and
as we shall see in the next section.

\section{Internal Heat Bath}
\label{cpld-subs}

In this section we remove the external heat bath and group all
the particles of the system into small subsystems labeled $A_{\mu}$.
Focussing our attention on one subsystem we can think of the
all the remaining ones as constituting a heat bath.  There will
be forces acting among these subsystems as sketched in Figure 2.
These forces can be separated into a part which is rapidly
varying on the time scale of interest and a part which is
slowly varying.

\begin{figure}[htb]
\setlength{\unitlength}{0.5mm}
\begin{center}
\begin{picture}(300,130)(0,0)
%
%    Bottom unit fig. 1-2
%
\thicklines
%
% tails length 5
\put(  5,100){\line( 1, 0){ 5}}
\put(290,100){\line( 1, 0){ 5}}
% 5 box labels
\put( 17,95){\LARGE $A_{\mu-2}$}
\put( 77,95){\LARGE $A_{\mu-1}$}
\put(142,95){\LARGE $A_{\mu  }$}
\put(197,95){\LARGE $A_{\mu+1}$}
\put(257,95){\LARGE $A_{\mu+2}$}
%
% box 40x30  (tot. width 40) 5 of these
\multiput(0,0)(60,0){5}{
\put(10,85){\line(1,0){40}} \put(10,115){\line(1,0){40}}
\put(10,85){\line(0,1){30}} \put(50, 85){\line(0,1){30}}
}
%
% spring 12x18 with tails 4-4 (tot. width 20) 4 of these
\multiput(0,0)(60,0){4}{
\put(50,  100){\line(1, 0){4.0}} \put(54.0,100){\line(1,6){1.5}}
\put(55.5,109){\line(1,-6){3.0}} \put(58.5, 91){\line(1,6){3.0}}
\put(61.5,109){\line(1,-6){3.0}} \put(64.5, 91){\line(1,6){1.5}}
\put(66.0,100){\line(1, 0){4.0}}
}
%
%
%       Top Unit Fig. 2
%
\thinlines
\put(120,115){\oval( 60,35)[t]}\put(180,115){\oval( 60,35)[t]}
\put( 90,115){\oval(120,45)[t]}\put(210,115){\oval(120,45)[t]}
\end{picture}
\end{center}
\vspace*{-4.5cm}
\caption[]{ A series of small subsystems, $A_{\mu{-}1}$, $A_{\mu}$,
$A_{\mu{+}1}$,  ..., interacting with
each other via random thermal forces (thin lines). The random
forces are classified into nearest neighbour $F^{(1)}_{\mu}(t)$,
next nearest neighbour $F^{(2)}_{\mu}(t)$, ... , and so on.
In addition each subsystem interacts with its nearest neighbours
via conservative forces (heavy lines) too.}
\label{fig:f_int}
\setlength{\unitlength}{1mm}
\end{figure}
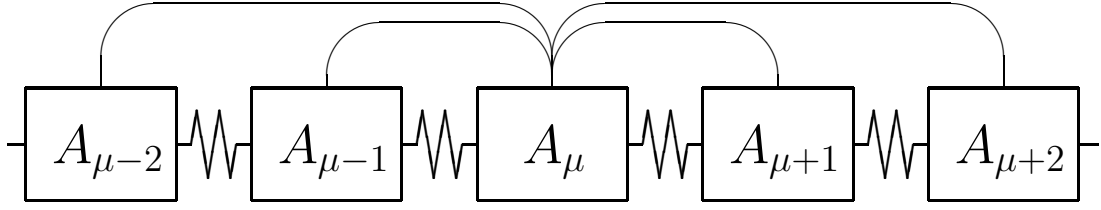

First consider a one dimensional system.  During a particular time
interval of duration $\tau$ there will be a net momentum transfer
$p_{\mu{-}1,\mu}$ from $A_{\mu{-}1}$ to $A_{\mu}$ and a net
momentum transfer $p_{\mu,\mu{+}1}$ from $A_{\mu}$ to $A_{\mu{+}1}$
due to the random forces.  The net force experienced by $A_{\mu}$ due
to these is
\be
F_{\mu}^{(1)}(t) = \frac{p_{\mu{-}1,\mu}-p_{\mu,\mu{+}1}}{\tau} \, .
\ee
The averaged equation of motion is just like eq. (14) except that
now the force $F$ originates in the intersubsystem interactions,
not with an external heat bath.  We make the assumption
that there are no correlations between random momentum transfers
involving different pairs of subsystems.  That is,
\be
\langle p_{\mu,\mu+1}(t) p_{\nu,\nu+1}(t-s) \rangle_0 =
T \tau^2 K^{(1)}(s) \delta_{\mu,\nu}
\ee
where as before, $K^{(1)}(s)$ is expected to fall exponentially with $s$.
Substitution into the equation of motion gives the following.
\be
m \langle v_{\mu}(t{+}\tau) - v_{\mu}(t) \rangle = G_{\mu} \tau
+ \int_t^{t{+}\tau} dt' \int^{t'-t}_0  ds \
\Big( v_{\mu{+}1}(t'{-}s) -2 v_{\mu}(t'{-}s) + v_{\mu{-}1}(t'{-}s)
\Big) K^{(1)}(s) \, .
\ee
Here each subsystem has been assumed to have the same mass, $m$, for
simplicity.

Next we should allow for the possibility of random forces acting
between next to nearest neighbors.  The arguments exactly parallel
those for nearest neighbors.  The force is
\be
F_{\mu}^{(2)}(t) = \frac{p_{\mu{-}2,\mu}-p_{\mu,\mu{+}2}}{\tau} \, .
\ee
The correlation functions are
\be
\langle p_{\mu,\mu+2}(t) p_{\nu,\nu+2}(t-s) \rangle_0 =
T \tau^2 K^{(2)}(s) \delta_{\mu,\nu}
\ee
This may be continued for next to next to nearest neighbors
{\it ad infinitum}.  The averaged equation of motion, taking all of them
into account, is
\begin{eqnarray}
\lefteqn{
m \langle v_{\mu}(t{+}\tau) - v_{\mu}(t) \rangle = G_{\mu} \tau
} %lefteqn
& &
\non
& & \qquad{}
+ \int_t^{t{+}\tau} dt' \int^{t'-t}_0  ds \,
\sum_{\sigma = 1}^{\infty}
\Big( v_{\mu{+}\sigma}(t'{-}s) -2 v_{\mu}(t'{-}s) +
 v_{\mu{-}\sigma}(t'{-}s) \Big) K^{(\sigma)}(s) \, .
\end{eqnarray}
The function $K^{(\sigma)}(s)$ undoubtedly decreases rapidly with
increasing $\sigma$.

In taking the continuum limit we replace the subsystem labels
$\mu$ and $\sigma$ with position coordinates $x$ and $y$, respectively.
We divide the equation by the average length, $l_0$,
of each subsystem.  The sum over $\sigma$ gets replaced by an
integral over $y$.
\begin{eqnarray}
\lefteqn{
\rho \, \langle v(x,t{+}\tau) - v(x,t) \rangle = g(x,t) \tau
} %lefteqn
& &
\non
& & \quad {}
+
\int_t^{t{+}\tau} dt' \int^{t'-t}_0  ds \int_0^{\infty}
dy \ k(y,s)\,
\Big( v(x{+}y,t'{-}s) -2 v(x,t'{-}s) + v(x{-}y,t'{-}s) \Big) \, .
\end{eqnarray}
Here $k(y,s)$ is the continuation of $K^{(\sigma)}(s)/l_0^2$.
Generally, the velocity will be more slowly varying with position
than the correlation function characterizing the random forces.
If that is so then the difference
$v(x+y,t'{-}s) -2 v(x,t'{-}s) + v(x-y,t'{-}s)$
is well approximated by $y^2 \partial_x^2 v(x,t'{-}s)$.  Then
\be
\rho \, \langle v(x,t{+}\tau) - v(x,t) \rangle = g(x,t) \tau
+ \int_t^{t{+}\tau} dt' \int^{t'-t}_0  ds \ \partial_x^2 v(x,t'{-}s)
\int_0^{\infty} dy \ y^2 \ k(y,s) \, .
\ee

To obtain a Langevin equation we proceed as before.  Divide through
by $\tau$, replace $t'$ with $t$ in the argument of the velocity
on the right side, let the upper limit on the $s$ integration go to
infinity, and denote the displacement by the variable $\eta(x,t)$.
\be
\rho \, \ddot{\eta}(x,t) = g(x,t) + \int_0^{\infty} ds \
\partial_x^2 \dot{\eta}(x,t{-}s)
\int_0^{\infty} dy \ y^2 \ k(y,s)
+ f'(x,t) \, .
\ee
The rapidly fluctuating force per unit length, $f'$, is the deviation from
the average value in the perturbed system.  It is constructed by
taking the continuum limit of the sum of random forces acting
on the subsystem minus the average of those forces, similar to eq. (19).
The average is just the second term on the right side of the above
equation.
\be
\frac{1}{l_0} \sum_{\sigma = 1}^{\infty} F_{\mu}^{(\sigma)}(t)
- \langle f(x,t) \rangle \rightarrow f'(x,t)
\ee
In general it is difficult to find a simple closed expression for $f'$.
One way is to generate it from knowledge of the
correlation functions like eqs. (29) and (32).

When our interest is on times and lengths greater than those
characterizing the friction kernel and when the actual displacement
velocity is slowly varying on those scales we can take the
velocity outside the integration to obtain
\be
\rho \, \ddot{\eta}(x,t) = g(x,t) + \gamma^* \ \partial_x^2 \dot{\eta}(x,t)
+ f'(x,t)
\ee
where
\be
\gamma^* = \int_0^{\infty} ds \int_0^{\infty} dy \ y^2 \ k(y,s) \, .
\ee
The essential difference between this equation and the one for
an external heat bath is the second space derivative acting on
the velocity, which is absent in the latter case.  The origin
is translational invariance.  The external heat bath imposes
a particular frame of reference, whereas the absence of an
external heat bath means that the system must be invariant
under boosts of constant velocity.

To generalize to more than one spatial dimension is straightforward.
We assume that random momentum transfers between different pairs
of subsystems are uncorrelated and that the system is rotationally
invariant, for simplicity.
\be
\langle
p^i_{\vec{\mu},\vec{\mu}+\vec{\sigma}}(t')
p^j_{\vec{\nu},\vec{\nu}+\vec{\rho}}(t'-s) \rangle_0
=
T \tau^2
\delta_{\vec{\mu},\vec{\nu}}\delta_{\vec{\sigma},\vec{\rho}} \,
K^{(\vec{\sigma})}_{ij}(s)
\ee
This can be used to calculate the force-force correlation function.
\bey
\beta \langle F_{\vec{\mu}}^i(t') F_{\vec{\nu}}^j(t'-s) \rangle_0
& = &
\beta \sum_{\vec{\sigma},\vec{\rho}}
\langle F_{\vec{\mu}}^{(\vec{\sigma})i}(t')
F_{\vec{\nu}}^{(\vec{\rho})j}(t'-s) \rangle_0
\non
& = &
\sum_{\vec{\sigma}}
\Big(
2 \delta_{\vec{\mu},\vec{\nu}}
- \delta_{\vec{\mu}+\vec{\sigma},\vec{\nu}}
- \delta_{\vec{\mu}-\vec{\sigma},\vec{\nu}}
\Big)
K^{(\vec{\sigma})}_{ij}(s)
\eey
The sum over $\mbf{\sigma}$ runs from $\sigma_1, \sigma_2, \sigma_3
= 1$ to $\infty$, and similarly for $\mbf{\rho}$, in order not to double
count the number of pairs.  Then
\be
- \beta \langle F_{\vec{\mu}}^i(t') F_{\vec{\nu}}^j(t'-s) \rangle_0
v_{\vec{\nu}}^j(t'-s) = \sum_{\vec{\sigma}}
\left( v^j_{\vec{\mu}{+}\vec{\sigma}}(t'{-}s) -2 v^j_{\vec{\mu}}(t'{-}s) +
 v^j_{\vec{\mu}{-}\vec{\sigma}}(t'{-}s) \right)
K^{(\vec{\sigma})}_{ij}(s) \, .
\ee
This leads to an equation of motion similar to eq. (33) with
spatial indices $i$ and $j$ in the appropriate places
and with the scalars $\mu$ and $\sigma$ replaced with location vectors
$\mbf{\mu}$ and $\mbf{\sigma}$.

We take the continuum limit in the usual way.
\begin{eqnarray}
\lefteqn{
\rho \, \langle v^i({\bf x},t{+}\tau) - v^i({\bf x},t) \rangle
=
g^i({\bf x},t) \tau
} %lefteqn
& &
\non
& & {}
+\int_t^{t{+}\tau} dt' \int^{t'-t}_0  ds
\int_0^{\infty} dy_1 dy_2 dy_3 \ k^{ij}({\bf y},s)
\left( v^j({\bf x}{+}{\bf y},t'{-}s) -2 v^j({\bf x},t'{-}s)
+ v^j({\bf x}{-}{\bf y},t'{-}s) \right)  .
\non
\end{eqnarray}
Because of rotational symmetry the kernel $k$ has the same
structure as in eq. (23).  Expand
the velocity to second order in ${\bf y}$ and integrate over
all directions of ${\bf y}$.
\begin{eqnarray}
\lefteqn{
\rho \, \langle v^i({\bf x},t{+}\tau) - v^i({\bf x},t) \rangle
=
g^i({\bf x},t) \tau
} %lefteqn
& &
\non
& & \qquad {}
+{1\over 120}
\int_t^{t{+}\tau} dt' \int^{t'-t}_0  ds \,
\nabla^2 v^i({\bf x},t'-s) \,
\int d^3y \ y^2 \,
\Big( k_{\rm L}(y,s) + 4 k_{\rm T}(y,s) \Big)
\non
& & \qquad {}
+{1\over 60}
\int_t^{t{+}\tau} dt' \int^{t'-t}_0  ds \,
\partial_i \nabla{\cdot}{\bf v}({\bf x},t'-s)
\int d^3y \ y^2 \,
\Big( k_{\rm L}(y,s) - k_{\rm T}(y,s) \Big)
\end{eqnarray}
Here $y \equiv |{\bf y}|$.  Converting this to a Langevin equation
in the now familiar way we get
\begin{eqnarray}
\lefteqn{
\rho \, \ddot{\mbox{\boldmath$\eta$}}({\bf x},t)
= {\bf g}({\bf x},t)
+
\frac{1}{60}
\int_0^{\infty} ds \,
\nabla (\nabla{\cdot}\dot{\mbox{\boldmath$\eta$}}({\bf x},t{-}s))
\int d^3y \ y^2 \
\Big( k_{\rm L}(y,s) - k_{\rm T}(y,s) \Big)
}% lefteqn
\non
& & \quad {}
+
\frac{1}{120}
\int_0^{\infty} ds \,
\nabla^2 \dot{\mbox{\boldmath$\eta$}}({\bf x},t{-}s)
\int d^3y \ y^2 \
\Big( k_{\rm L}(y,s) + 4 k_{\rm T}(y,s) \Big)
+
{\bf f}^{\prime}({\bf x},t) \, .
\end{eqnarray}
The rapidly fluctuating force ${\bf f}'$ may be constructed along
the same lines as in one dimension.

In the very low frequency limit the velocities can be evaluated
at time $t$ and taken past the $s$ integral.  The Langevin
equation then reduces to the Navier-Stokes equation
\be
\rho \frac{d{\bf v}}{dt} = -\nabla P + \eta \nabla^2 {\bf v}
+ \left( \zeta + \frac{1}{3}\eta \right) \nabla 
\left( \nabla \cdot {\bf v} \right)
\ee
with the addition of the rapidly fluctuating force.  By comparison we
can determine the shear ($\eta$) and bulk ($\zeta$) viscosities.
\begin{eqnarray}
\eta &=& \frac{1}{120} \int_0^{\infty} ds \int d^3y \ y^2
\ \left( 4k_{\rm T} + k_{\rm L} \right) \nonumber \\
\zeta &=& \frac{1}{72} \int_0^{\infty} ds \int d^3y \ y^2
\ \left( k_{\rm L} -2k_{\rm T} \right)
\end{eqnarray}
The requirement that the viscosities must be nonnegative places
a restriction on the relative magnitudes of the moments
of the longitudinal and transverse correlation functions.

\section{Comparison of Field Equations and Conclusion}

An isotropic gas or liquid can be fully described by a scalar density
$\phi \equiv - \nabla{\cdot}\mbf{\eta}$ characterizing the local relative
compression of the matter.  The Langevin equation for the cases
of external and of internal heat baths can be expressed in terms
of $\phi$ alone without any reference to a vector displacement.
But first we should inquire about the effective Lagrangian describing
this field in the absence of dissipation and fluctuation.

A real scalar field has the Lagrangian density
\be
{\cal L}_0 =
\frac{1}{2} \dot{\phi}^2 - \frac{c^2}{2} (\nabla \phi)^2
- V(\phi)
\ee
where $V(\phi)$ is a potential, usually a polynomial.  This
Lagrangian density does not contain the fluctuations nor any
dissipative term.  For example, if $V = 0$ then this describes
undamped sound waves with speed $c$.  The equation of motion is
\be
\ddot{\phi} = c^2 \nabla^2 \phi -\partial V(\phi)/\partial \phi \, .
\ee
Here we may identify $\nabla{\cdot}{\bf g}$ with
$\partial V(\phi)/\partial \phi$.
The influence of fluctuations can be included by adding to the
Lagrangian density
\be
{\cal L}_{\rm fluc} = - b'({\bf x},t) \phi \, .
\ee
The fluctuating field $b'$ is simply identified with
$-\nabla{\cdot}{\bf f}'/\rho$
in both the external and internal heat bath systems.
The nature of this stochastic field is determined by the microscopic
dynamics of the system expressed in terms of its original degrees
of freedom.  The frictional term cannot, of course, be written
as an additional term in the Lagrangian.  It must be added by
hand to the equation of motion.

Including the dissipative and the fluctuating terms the
Langevin field equation for an external heat bath is
\be
\ddot{\phi}({\bf x},t) = c^2 \nabla^2 \phi({\bf x},t)
 -\partial V(\phi)/\partial \phi
- \gamma \dot{\phi}({\bf x},t) + b'({\bf x},t)
\ee
where $\gamma$ was discussed in section 2.3.  For the internal
heat bath
\be
\ddot{\phi}({\bf x},t) = c^2 \nabla^2 \phi({\bf x},t)
 -\partial V(\phi)/\partial \phi
+ \gamma^* \nabla^2 \dot{\phi}({\bf x},t) + b'({\bf x},t)
\ee
where
\be
\gamma^* =
\frac{1}{120 \rho} \int_0^{\infty} ds \int d^3y
\ y^2 \
\Big( 3 k_{\rm L}(y,s) + 2 k_{\rm T}(y,s) \Big)
\, .
\ee
This is obtained under the assumption that the variation in
$\phi$ is minor during the correlation times of
$k_{\rm L}$ and $k_{\rm T}$.

It is important to appreciate the difference between the external
and internal heat baths.  With an external heat bath there is a
preferred frame of reference.  The random forces couple individual
Brownian particles to the heat bath; it is assumed that there are
no random forces between Brownian particles.  With an internal
heat bath there is no preferred frame of reference.  The random
forces couple different subsystems.  This difference is the origin
of the Laplacian in eq. (52) which is what really distinquishes
the two field equations.  In fact, if we desired a more fine-grained
description we should expand the velocities in eqs. (15) and (43)
to higher order in spatial and temporal variations.  The resulting
field equation will involve dissipation of the form
\be
\sum_{i,j = 1}^{\infty} \Gamma_{ij} \left( \nabla^2 \right)^{i-1}
\left(\frac{\partial}{\partial t} \right)^j \ \phi({\bf x},t) \, .
\ee
Translational invariance demands that $\Gamma_{1j} = 0$.  All
the coefficients can be determined in terms of moments of the
friction kernel.  Generally we would expect that only the first
few terms in the sum are required for a good description of a
coarse-grained system.

Of course, in realistic situations it is always a matter of judgement
whether there is a clear case of external or internal heat bath. It is
possible to have different components in the system,
where part of this system can be considered as internal while the
rest is external.  Usually the external heat bath is chosen to be
dominant, but this is not necessarily the case and mixed cases
may come up in realistic studies.

Although it is not our intention to apply these results to any
particular problems in this paper it is instructive to consider
one example.  Let $V(\phi) = 0$ and look for plane wave solutions
in the absence of significant fluctuations.  Denoting the frequency
of the wave by $\omega$ and the wave vector by $q$ we obtain the
dispersion relation
\be
\omega = \sqrt{c^2 q^2 - \gamma^2/4} -i\gamma/2
\ee
for the external heat bath, and
\be
\omega = \sqrt{c^2 q^2 - \gamma^{*\,2}q^4/4} -i\gamma^* q^2/2
\ee
for the internal heat bath.  For a real wave vector the frequency,
in each case, has an imaginary part resulting in damping of the wave.
Note that for the external heat bath the wave becomes overdamped
for wave vectors less than $\gamma/2c$.  This is certainly not
representative of sound waves in the atmosphere!  For the
internal heat bath the damping goes to zero as $q^2$ resulting
in the absence of dissipation in this limit so that
$\omega \rightarrow cq$.  This dispersion relation is the
same as obtained for damped sound waves in a gas or liquid by
solution of viscous hydrodynamics \cite{Landau}.

In summary, we have derived (in some sense) coarse-grained
Langevin-type effective field equations based on the classical
dynamics of systems of many particles.  We considered two
extreme limits: one where the stochastic forces arose from
coupling to an external heat bath, the other where the stochastic
forces arose from statistical fluctuations in small parts of
the full system.  The obvious next steps are to consider
quantum mechanical effects, and to apply these results to
real atomic, molecular, or nuclear systems.
equations

\section*{Acknowledgements}

This work was supported by the US Department of Energy under grant
DE-FG02-87ER40328 and by the US National Science Foundation
under grant NSF/INT-9602108.

\end{document}